\documentclass[useAMS,usenatbib]{mnras}

\usepackage{graphicx}

\title[The Ca {\sc ii} triplet in red giant spectra]
{The Ca {\sc ii} triplet in red giant spectra: [Fe/H] determinations and the role of [Ca/Fe]}
\author[Da Costa]
{G. S. Da Costa$^1$\\
$^1$Research School of Astronomy and Astrophysics, Australian National 
University, Canberra, ACT 0200, Australia}
    
\begin{document}

\maketitle

\begin{abstract}

Measurements are presented and analyzed of the strength of the Ca {\sc ii} triplet lines in red giants in Galactic 
globular and open clusters, and in a sample of red giants in the LMC disk that have significantly different [Ca/Fe] 
abundance ratios to the Galactic objects.  The Galactic objects are used to generate a calibration
between Ca~{\sc ii} triplet line strength and [Fe/H], which is then used to estimate [Fe/H]$_{CaT}$ for the 
LMC stars.  The values are then compared with the [Fe/H]$_{spec}$ determinations from high dispersion 
spectroscopy.   After allowance for a small systematic offset the two abundance determinations are in excellent 
agreement.  Further, as found in earlier studies \citep[e.g.,][]{Ba08}, the difference is only a very weak function of 
the [Ca/Fe] ratio.  For example, changing [Ca/Fe] from +0.3 to --0.2 causes the Ca {\sc ii} based abundance to 
underestimate [Fe/H]$_{spec}$ by only $\sim$0.15 dex, assuming a Galactic calibration.  Consequently, the 
Ca {\sc ii} triplet approach to metallicity determinations can be used without significant bias to study stellar systems 
that have substantially different chemical evolution histories. 

\end{abstract}


\section{Introduction} \label{Intro} 

The lines of the Ca {\sc ii} triplet at $\lambda$8498, 8542, and 8662\AA\/ are the strongest lines in the far-red
spectral region in late-type red giant stars.  As a result, measurements of the strength of these lines have been 
extensively used as a metallicity indicator in a variety of stellar systems.   These include Galactic globular clusters
\citep[e.g.,][]{AD91, DA95, Ru97,Sa12}, Galactic open clusters \citep[e.g.,][]{AC04, WC09, Car09, Car15} and stars in 
the Galactic Bulge \citep[e.g.,][]{SV15}.  It has also been used to study both field stars and star clusters
in the LMC \citep[e.g.,][]{Olz91, AC05, Car08a} and SMC \citep[e.g.,][]{DH98, Car08b, Par10, Dob14, Par15}, 
as well as red giants in dwarf spheroidal \citep[e.g.,][]{AD91, Po04,  To04, Ba08, Ba11} and dwarf irregular galaxies 
\citep[e.g.,][]{RL13}.

The calibration of the measured Ca {\sc ii} line strengths is usually carried out by observing with the same 
instrument stars in systems whose abundances are known from another source.  Usually, but not always,
the calibration is carried out in terms of [Fe/H], which is adopted as a proxy for overall abundance.  
The alternative is to calibrate to [Ca/H]  \citep[e.g.,][]{JEN96, Bo07} although naturally this requires 
knowledge of the [Ca/H] values for the calibration objects.  The original studies concentrated primarily on metallicity 
determinations for relatively old and metal-poor systems.  Subsequent work, however, has demonstrated that the 
influence of age on the technique is relatively small \citep{AC04, Car07}, so that it can be applied to stellar systems 
with ages as young as $\sim$0.3 Gyr.   The form of the metallicity calibration has also been investigated at
both the metal-rich and metal-poor regimes.  At higher metallicities, i.e., near solar and above, the form of the 
line strength calibration relation remains relatively linear \citep[e.g.,][]{SV15}, but as shown by \citet{ES10}
and \citet{Car13},
at metallicities below [Fe/H] $\approx$ --2.5 the relation between Ca {\sc ii} triplet line strength and abundance
becomes notably non-linear.

A question that remains relatively little explored, however, is the role of [Ca/Fe] -- if the objects under study 
have very different 
[Ca/Fe] ratios from those in the calibrating objects, are the derived [Fe/H] values significantly biased? 
The answer to this 
question is important because, for example, low mass dwarf galaxies have a different chemical enrichment
history to the halo and disk of the Galaxy, generally showing substantially lower [alpha/Fe] ratios compared to
Galactic stars of similar [Fe/H] \citep[e.g.,][]{THT09}.  

\citet{Ba08} investigated this question \citep[see also][]{Id97, Car07, ES10} and found the effect of different
[Ca/Fe] abundance ratios is relatively minor, but further studies of the issue are required.  That then is the purpose
of this paper -- to investigate the influence of [Ca/Fe] on [Fe/H] determinations from the Ca {\sc ii} method.  This is
achieved by developing a standard Ca {\sc ii} line strength calibration using a number of Galactic calibrators,
and then applying the calibration to Ca {\sc ii} line strength measurements for red giants in the disk of the LMC\@.
The LMC stars have been observed at high resolution so that both [Fe/H] and [Ca/Fe] values are available,
allowing a direct comparison between the two abundance measures and an investigation of whether any 
differences are related to the [Ca/Fe] values.   
 
The observations and data reduction techniques are discussed in the following section, along with an 
evaluation of the membership status of the cluster stars observed.  In \S 3 the line strength analysis method adopted
is presented, while in \S 4 the abundance calibration relation is derived.  The results are then presented and 
discussed in the final section.

\section{Observations and Reductions}

The data for this study consist of observations of Galactic globular and open clusters,
which are used to provide a calibration between Ca {\sc ii} triplet line strength and abundance [Fe/H], and observations
of a sample of red giants in the disk of the LMC orginally studied at high dispersion by \citet{Po08}, and for which
an updated analysis is available in \citet{vdS13}.  All observations were carried out at the Anglo-Australian 
Telescope using the 2dF multi-object fibre positioner and the AAOmega dual beam spectrograph\footnote{Manuals
and technical information at http://www.aao.gov.au/science/instruments/current/AAOmega.} \citep{Sa04, Sh06}.
The red arm of the spectrograph was configured with the 1700D grating centred at $\lambda$8600\AA\/ to give
coverage of the Ca {\sc ii} triplet lines with a resolution {\it R} 
of approximately 10000.  The corresponding blue spectra are not used in this work.

Five Galactic globular clusters (NGC~104 (47~Tuc), 288, 1904, 2298 and 7099 (M30)) and two Galactic 
open clusters (Melotte 66 and M67) were observed over a number of different
nights as parts of other programs, usually when conditions were not optimal for the main program.  The LMC disk
field was observed as part of a larger program concentrating on the LMC\@.
Details of the observations of the Galactic calibrators and the LMC disk field are given Table \ref{Tab1}.

For the globular clusters the input catalogues to the fibre allocation software
were drawn initially from the position and photometry information available 
in the photometric standard star fields database maintained by 
Stetson\footnote{http://www3.cadc-ccda.hia-iha.nrc-cnrc.gc.ca/community/STETSON/standards/}.  
Stars brighter than the horizontal branch on both the red giant (RGB) and asymptotic 
giant branches (AGB) were selected with final positions sourced from the UCAC4 catalogue \citep{Za13}.  
For Melotte~66 candidate red giants at the magnitude of the red clump and brighter were selected 
from the photometry of \cite{AT94} with positions again sourced from the UCAC4 catalogue.
The Stetson photometry and position information were also used as the initial source for the open cluster M67.
Here, given the sparseness of the giant branch, the input catalogue included 3 red clump stars as well as 2
additional bright red giant members that lack Stetson photometry.  Positions were again taken from the UCAC4 
catalogue.

\begin{table}
\caption{Observations}
\label{Tab1}
\begin{tabular}{lcl}
\hline
Target & Date & Exposure (sec) \\ 
\hline
NGC 104 (47 Tuc) & 2011 Oct 30  & 3 $\times$ 600\\
NGC 288 & 2011 Oct 31 & 3 $\times$ 1200\\
NGC 1904 & 2011 Oct 31 & 3 $\times$ 1000\\
NGC 2298 & 2011 Oct 30 & 3 $\times$ 1000\\
NGC 7099 (M30) & 2011 Oct 31 & 2 $\times$ 1200\\
 & & \\
Melotte 66 & 2011 Nov 01 & 3 $\times$ 1000\\
M67 & 2010 Dec 29 & 3 $\times$ 300\\
 & & \\
LMC Disk & 2010 Dec 28 & 2 $\times$ 1800\\ 
\hline
\end{tabular}
\end{table}

\begin{figure}
\centering
\includegraphics[angle=-90.,width=0.46\textwidth]{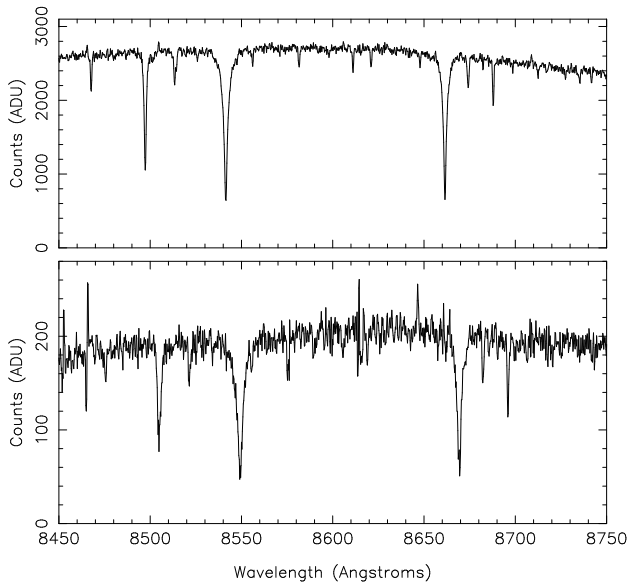}
\caption{Typical final reduced spectra.  The upper spectrum is of the red giant S752 in the globular cluster
NGC~288, which has [Fe/H] = --1.32 and [Ca/Fe] = 0.41.  The star has $V$ = 13.97.
The lower spectrum is of the LMC red giant 0664-LMCDisc01 for which \citet{vdS13} list [Fe/H] = --0.49 
and [Ca/Fe] = --0.17.  The star has $V$ = 17.16.   
Neither spectrum has been corrected to rest wavelength. \label{fig1}}
\end{figure}

The observations were reduced with the 2dF data reduction pipeline {\sc 2dfdr}\footnote{http://www.aao.gov.au/science/software/2dfdr} using the standard approach in which
fibre flat-field exposures set the location of the fibres on the detector and arc-lamp exposures provide the
wavelength calibration.  Sky subtraction was achieved using the {\sc skyflux(med)} approach, in which the relative throughputs of the fibres are determined from the observed intensities of night-sky emission lines present in
the raw spectra.  The wavelength-calibrated sky-subtracted spectra from the individual integrations were then 
combined to remove cosmic ray contamination.  Final signal-to-noise ratios were generally excellent 
($>$50 pixel$^{-1}$) for 
the spectra of the Galactic objects but are much lower ($\sim$15--20 pixel$^{-1}$) for the LMC disk stars because of 
the fainter magnitudes and poorer seeing (FWHM $\sim$ 2$^{\prime\prime}$) during the observations.  
Fig.\ \ref{fig1} shows two example spectra -- a red giant in the globular cluster NGC~288 and a LMC disk star.

\subsection{Globular Cluster Membership}

Two approaches were used to determine the membership status for the stars observed in the Galactic globular
clusters.  The first involved measuring the radial velocities of the stars to detect any radial velocity non-members.  
The radial velocities were determined from the observed wavelengths of the Ca {\sc ii} triplet lines using the 
IRAF\footnote{Information and distribution for IRAF is available through http://iraf.noao.edu/.} routine {\sc rvidlines}.
The observed velocities were corrected to heliocentric values and then compared to the mean cluster velocities
tabulated in the 2010 on-line version\footnote{http://physwww.mcmaster.ca/$\sim$harris/mwgc.dat} 
of the \citet{H96} catalogue\footnote{Hereafter all references to the \citet{H96} catalogue refer to the 2010
on-line version.} and likely non-members flagged.  
The precision of radial velocities observed with this instrumental setup is 2 km s$^{-1}$ or better for spectra with
good signal-to-noise \citep[e.g.,][]{GDC12} making membership discrimination straightforward.  For the 5 globular
clusters the numbers of radial velocity non-members in the observed samples were 2 (47~Tuc), 
4 (NGC~288), 8 (NGC~1904), 8 (NGC~2298) and 1 (NGC~7099).  

The second approach relies on the reasonable assumption that the stars in each cluster have approximately 
the same overall abundance so
that there should be a monotonic relationship between the measured strength of the Ca {\sc ii} lines and a luminosity
indicator such as ($V-V_{HB}$), where $V_{HB}$ is the apparent magnitude of the horizontal branch as given in
the \citet{H96} catalogue.  The measurement of the Ca {\sc ii} line strengths is described in the following section,
here we are only interested in stars that do not follow the relation exhibited by the majority of the cluster stars.
Such stars are classified as line strength non-members.  For the cluster samples observed here only three 47~Tuc
stars were classified as line strength non-members.  The final mean cluster radial velocities and the number of 
members are given
in Table \ref{Tab2}.  The agreement with the velocities tabulated by \cite{H96} is excellent -- the largest difference
is for NGC~2298 where the value determined here is 146.2 $\pm$ 0.6 km s$^{-1}$ while \citet{H96} lists 
148.9 $\pm$ 1.2 km s$^{-1}$.

\begin{table}
\begin{minipage}{0.98\textwidth}
\caption{Radial Velocities}
\label{Tab2}
\begin{tabular}{lcr}
\hline
Target & N$^a$ & Radial Velocity \\
 & & km s$^{-1}$ \\
\hline
NGC 104 (47 Tuc) &  64 & --18.2 $\pm$ 0.9\\
NGC 288 &  40 & --45.1 $\pm$ 0.4 \\
NGC 1904 & 22 & 205.9 $\pm$ 0.7\\
NGC 2298 & 12 & 146.2 $\pm$ 0.6\\
NGC 7099 (M30) & 25 & --184.3 $\pm$ 0.5\\
 & & \\
Melotte 66 & 17 & 19.8 $\pm$ 0.9\\
M67 & 10 & 32.5 $\pm$ 0.5 \\
% & & \\
%LMC Disk & 2010 Dec 28 & 2 $\times$ 1800\\ 
\hline
\end{tabular}

$^a$ Number of cluster members, including both RGB and AGB \\
stars for 47 Tuc, NGC 288 and NGC 1904.
\end{minipage}

\end{table}

\begin{figure}
\centering
\includegraphics[angle=-90.,width=0.46\textwidth]{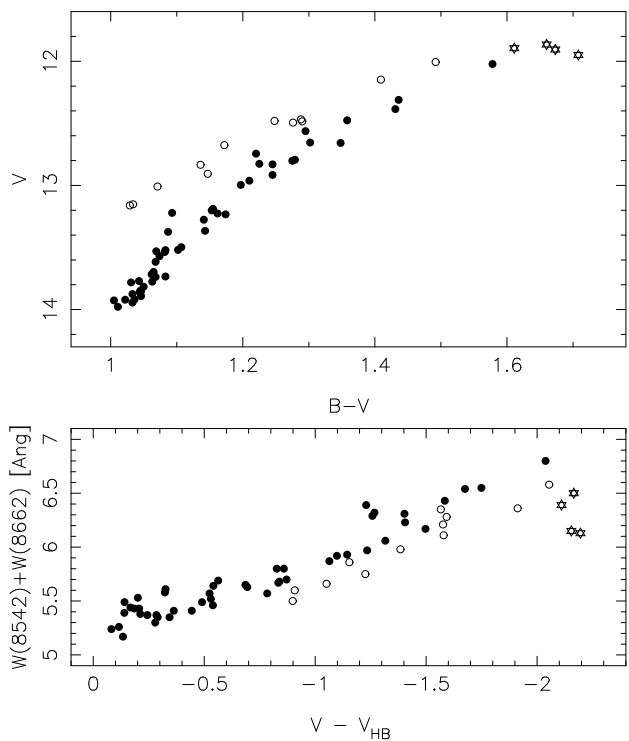}
\caption{The top panel shows the CMD for the 47~Tuc cluster members using photometry from the Stetson
database.  Filled symbols are RGB stars, open symbols are AGB stars and the star symbols are the four
most luminous giants in the sample.  The lower panel shows the sum of the strengths of the $\lambda$8542\AA\/ 
and $\lambda$8662\AA\/ lines of the Ca {\sc ii} triplet against $V-V_{HB}$ for the same stars.  The AGB stars
have systematically lower line strengths at a given $V-V_{HB}$.  The four most luminous stars have apparently
weaker line strengths because the onset of TiO bands in their spectra.
\label{fig1a}}
\end{figure}

An additional pair of points can be made concerning the globular cluster samples.  First, TiO bands are clearly
present in the spectra of the most luminous stars in the cluster 47~Tuc.  The presence of TiO depresses the 
local continuum against which the line strengths are measured (see \S \ref{EWmeas}) leading to anomalously low
line strengths.  As a result the 4 brightest stars in the 47~Tuc sample, all of which have $V-V_{HB}$ $\leq$ --2.10,
have been excluded from further analysis.  The onset of the effects of TiO is, at least in $V-V_{HB}$, quite 
rapid as the next most luminous 47~Tuc red giant observed, which has  $V-V_{HB}$ = --2.04, is clearly unaffected.  
This is not surprising as the colour-magnitude diagram (CMD) for the cluster shows that the $V$ magnitudes are 
decreasing only slowly near the RGB-tip while the colours become substantially redder (see Fig.\ \ref{fig1a}).

The second point relates to the presence of AGB stars in the observed samples.  In the cluster CMDs, given the
exquisite precision of the Stetson photometry, AGB stars are readily separated from the RGB\@.  The AGB stars are
brighter than the RGB at the same colour, or equivalently at the same $V-V_{HB}$, the AGB stars are bluer (hotter). 
The effect is that at a given $V-V_{HB}$, the AGB stars have a lower line strength compared to the RGB stars.  
This is illustrated in Fig.\ \ref{fig1a} which shows the CMD for the 47~Tuc cluster members in the upper panel and
the (Ca {\sc ii} line strength, $V-V_{HB}$) diagram the lower panel.  The AGB stars lie systematically below the RGB
stars at the same $V-V_{HB}$ value, in this case by $\sim$0.16\AA.  Consequently, we have excluded 
the AGB stars observed from the 47~Tuc,
NGC~288 and NGC~1904 samples; no AGB stars were included in the NGC~2298 and M30 samples. 

\subsection{Melotte 66 and M67 membership}

For Melotte 66 a sample of 20 stars were observed.  Two stars were immediately identified as non-members
on the basis of discrepant radial velocities.  Nine of the remaining stars are classified as cluster members in
\citet{AC04}, a categorization consistent with the velocities and line strengths derived here.  As for the balance
of the sample, eight of the nine stars also have velocities and line strengths consistent with membership.  
The final star is Mel66-W1580 (WEBDA id number), which \citet{Ses08} indicate is 
a fast rotator.  The radial velocity determined here is 50.9 $\pm$ 2.0 km~$s^{-1}$ while that given
by \citet{Clb14} is --14 $\pm$ 17 km~$s^{-1}$; the star is probably a binary.  It has not been considered further.
The mean velocity of the 17 stars classified as members is 19.8 $\pm$ 0.9 km s$^{-1}$ (std error of mean) which
agrees well with the values of 20.7 $\pm$ 0.6 and 22.1 $\pm$ 0.9 km s$^{-1}$ given by \citet{Ses08}
and \citet{Clb14}, respectively.

The ten M67 stars observed are all known to be cluster members and the current data support that classification. 
The mean radial velocity for these stars is
32.5 $\pm$ 0.5 km s$^{-1}$ which is in excellent agreement with other determinations.
For example, \citet{Clb14} give the cluster radial velocity as 33.1 $\pm$ 0.6 km s$^{-1}$.  Six of the stars were
also observed by \citet{AC04}.

\subsection{The LMC disk stars}

The 2df configuration software allowed 57 of the 59 stars in the \citet{Po08} sample to be observed.  
Unfortunately a CCD bad column fell exactly on the $\lambda$8662\AA\/ line for star 1118-LMCDisc01 (using
the naming convention of \citet{vdS13} -- this is star RGB\_1118 in the \citet{Po08} list) and thus the spectrum
of this star could not be used in the analysis.  
The seeing during the observations varied from FWHM 1.5$^{\prime\prime}$ to
2.5$^{\prime\prime}$ leading to the possibility of contamination of the individual stellar spectra by neighbouring 
stars, given the 2dF fibre
diameter of $\sim$2$^{\prime\prime}$ and the crowded nature of the LMC disk field.  As a result, only those
stars whose observed radial velocity agreed well with that given by \citet{Po08} were retained.  There are
33 such stars and for this sample the mean difference in radial velocity ($v_{AAT}-v_{P08}$) is 0.4 km s$^{-1}$
with a standard deviation of 2.9 km s$^{-1}$.  A further cut on the sample was applied by requiring the 
signal-to-noise to exceed 15 per pixel
in the predominantly continuum region between the stronger pair of Ca {\sc ii} lines.  
This left a final sample of 32 LMC disk stars for analysis.

\section{Line Strength Analysis} \label{EWmeas}

The strengths of the $\lambda$8542\AA\/ and $\lambda$8662\AA\/ lines of the Ca~{\sc ii} triplet were measured 
on all the spectra using feature and continuum band passes identical to those specified originally in 
\citet{AD91}.  Unlike that work, however, which fitted gaussians to the line profiles, the psuedo equivalent widths 
were determined by fitting a gaussian plus lorentzian function using a procedure similar to that described in 
\citet{AC04}.  The equivalent widths were then summed to form the line strength index $\Sigma$W and the values
for the cluster member stars plotted against $V-V_{HB}$.  The $V_{HB}$ values for the globular clusters were
taken from the \citet{H96} catalogue while for Melotte 66 and M67, $V_{HB}$ was taken as the mean 
magnitude of the red clump in the CMDs using the photometry sources cited above.

As is now well-established \citep[e.g.,][]{AD91,Ru97,Car07,Sa12} for red giants brighter than approximately
the magnitude of the horizontal branch, and for the metallicity range covered by the calibration objects, the 
slope of the 
relation between $\Sigma$W and $V-V_{HB}$ is essentially independent of metallicity.  Consequently, the 
weighted average of the slopes of the least squares fits to the individual cluster data was calculated, and then
refit to the individual data sets.  The average slope found was --0.660 $\pm$ 0.016 \AA\/mag$^{-1}$ which is 
consistent with other determinations of this parameter; e.g., \citet{Sa12} found a value of --0.627 \AA\/mag$^{-1}$.

\begin{figure}
\centering
\includegraphics[angle=-90.,width=0.46\textwidth]{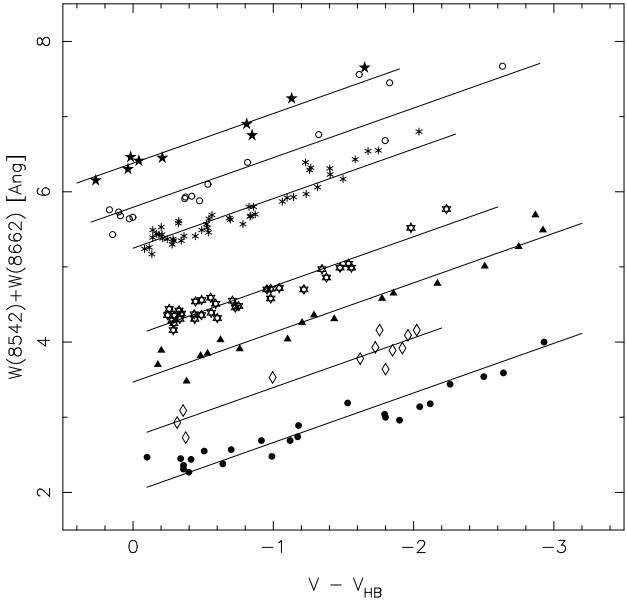}
\caption{The sum of the strengths of the $\lambda$8542\AA\/ and $\lambda$8662\AA\/ lines of the Ca II triplet 
in red giants in the Galactic calibration clusters is plotted magnitude difference from the horizontal 
branch ($V-V_{HB}$).  The clusters are NGC~7099 (filled circles), NGC~2298 (open diamonds), NGC~1904 (filled
triangles), NGC~288 (6-point stars), 47~Tuc (asterisks), Melotte~66 (open circles) and M67 (filled stars).  
The solid lines fitted to each data set have a slope of --0.66 \AA/mag. \label{fig2}}
\end{figure}

In Fig.\ \ref{fig2} the individual $\Sigma$W values are plotted against $V-V_{HB}$ for each of the Galactic calibration 
clusters.  The {\it rms} deviation about the fitted lines lie in the range 0.11 -- 0.15 \AA, which is consistent with the 
measurement errors in the $\Sigma$W values.  An independent check is provided by a 
comparison with the values, denoted by $\Sigma$W$_{Cole}$, given in \citet{AC04} for the 15 Melotte~66 
and M67 stars in common, bearing in mind that  \citet{AC04} used the sum of all three Ca~{\sc ii} lines rather than 
just the sum of the strongest two.  For 14 of the 15 stars there is an excellent correlation between the two data sets: 
\begin{equation}
\label{Coleeqn}
\Sigma W_{Cole} = 1.216 \pm 0.100 \Sigma W - 0.111 \pm 0.608
\end{equation}
The {\it rms} about the relation is 0.15\AA, which suggests the $\Sigma$W errors are of order 0.11\AA\/ in both
data sets, which have excellent signal-to-noise values.  
The one discrepant star is W2236 in Melotte~66 for which the listed $\Sigma$W$_{Cole}$ value is 
$\sim$0.7 \AA\/ ($\sim$9\%) weaker than the value predicted from the measurement here and equation 
\ref{Coleeqn}.  Since the value measured here is consistent with the other Melotte~66 stars in Fig.\ \ref{fig2} the
star has been retained.  The identifications, positions, $V$ magnitudes, W(8542) and W(8662) line strengths,
and heliocentric radial velocities for all the cluster stars contributing to Fig.\ \ref{fig2} are given in the 
associated on-line material.  Table \ref{tab3} outlines the format of the data file.

\begin{table*}
\begin{minipage}{1.5\textwidth}
\caption{Observational Data for Cluster Member Stars$^a$}
\label{tab3}
\begin{tabular}{lrrrccc}
\hline
Star ID & RA(J2000) & Dec(J2000) & $V$ & W(8542) & W(8662) & Radial Velocity \\
 & & & & (\AA) & (\AA) &km s$^{-1}$ \\
\hline
M67-IV-202 &  08 50 12.30 & +11 51 24.5  &  8.84 &   4.29 & 3.36  &  32.3\\
M67-T626   &  08 51 20.10 & +12 18 10.4  &  9.36  &  4.05 & 3.19  &  33.5\\
M67-F170   &  08 51 29.91 & +11 47 16.8  &  9.64   & 3.77 & 2.98  &  33.6\\
M67-F108   &  08 51 17.48 & +11 45 22.7  &  9.69  &  3.84  & 3.06 &   33.4\\
M67-F105   &  08 51 17.10 & +11 48 16.1  & 10.29  &  3.60 & 2.85 &   33.0\\
\hline
\end{tabular} 
\end{minipage}
$^a$ The full Table is available in the associated on-line material.  This version outlines the format of the 
data file.
\end{table*}

Because of the lack of dependence on metallicity of the slope of the relations shown in Fig.\ 
\ref{fig2}, the information for each cluster can be reduced to a single parameter, the reduced equivalent width
$W^{\prime}$, which is defined as the value of $\Sigma$W at $V-V_{HB}$ = 0.  We now explore the relation
between $W^{\prime}$ and [Fe/H] for these Galactic calibration objects, noting that the uncertainties in the 
$W^{\prime}$ values are small: they range between 0.02 and 0.05 \AA.

\section{Abundance Calibration}

\begin{figure}
\centering
\includegraphics[angle=-90.,width=0.46\textwidth]{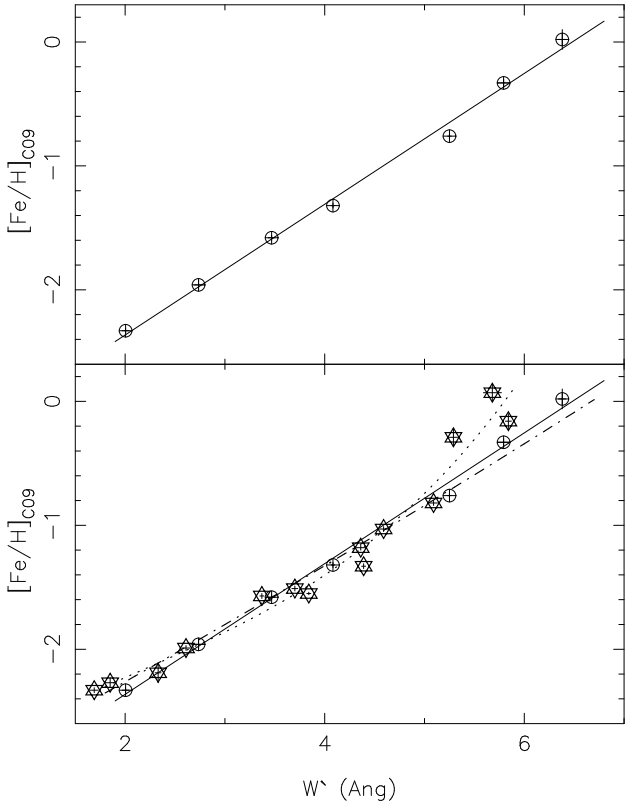}
\caption{{\it Upper panel:} The reduced equivalent width $W^{\prime}$ for each of the Galactic calibration objects 
observed here is 
plotted against [Fe/H].  In order of increasing [Fe/H] the objects are the globular clusters NGC~7099, NGC~2298, 
NGC~1904, NGC~288 and 47~Tuc and the open clusters Melotte 66 and M67.  The [Fe/H] values for the globular
clusters come from \citet{Car09} while those for Melotte 66 and M67 come from \citet{Ses08} and \citet{DY05},
respectively.  The solid line is a least squares fit to these data.  {\it Lower panel:}  The data from the upper panel are
augmented with the $W^{\prime}$ values for 14 additional standard globular clusters from \citet{Sa12} (six point star
symbols).  Shown also, as the dotted line, is the 3rd order calibration equation from \citet{Sa12} and, as the dot-dash
line, the 2nd order calibration relation from \citet{SV15}. \label{fig3}}
\end{figure}

In the upper panel of Fig.\ \ref{fig3} the value of  $W^{\prime}$ for each Galactic calibration object is plotted 
against the adopted [Fe/H] value.  For the globular clusters the [Fe/H] values are taken from Appendix I of 
\citet{Car09} where the
tabulated abundances can be traced back to a consistent set of high dispersion analyses of red giants in a
number of clusters \citep{Car09}.  For the open clusters Melotte 66 and M67 the results
of high dispersion studies have also been adopted: that of \cite{Ses08} who find [Fe/H] = \mbox{--0.33} $\pm$ 0.03 
for Melotte 66, and that of
\citet{DY05} who find [Fe/H] = +0.02 $\pm$ 0.08 for M67.  A linear least squares fit to the ($W^{\prime}$, [Fe/H]) 
data points is also shown in the Figure.  This fit is given by the equation:
\begin{equation}
\label{eq1}
[Fe/H] = 0.528 \pm0.017 W^{\prime} - 3.420 \pm 0.077
\end{equation}
with the {\it rms} about the fit being 0.06 dex.  There is no convincing indication of any deviation from linearity within 
the range for which the equation is valid: --2.4 $\la$ [Fe/H] $\la$ +0.1 dex, although as noted by \citet{ES10} and
\citet{Car13} a linear relation ceases to be appropriate in the extremely metal-poor regime.

It is of interest to compare this calibration with other recent determinations.  For example, \citet{Sa12} used 
essentially identical measurement and analysis techniques as those employed here to determine $W^{\prime}$
values for a number of Galactic globular clusters from samples of red giants observed with the FORS2 
multi-object spectrograph.  Although there are no clusters in common, the similarity in approach suggests that the
$W^{\prime}$ values for the calibration clusters in \citet{Sa12} should be directly analogous to those measured
here.  The lower panel of Fig.\ \ref{fig3} shows these $W^{\prime}$ values again plotted against [Fe/H] values from
\citet{Car09}.  \citet{Sa12} adopted a cubic calibration relation and that is also shown on the Figure.  In general
there is quite reasonable agreement between this work and that of \citet{Sa12} for the clusters less metal-rich than
[Fe/H] $\approx$ --0.8 dex (i.e., 47~Tuc and M71) but the more metal-rich clusters in \citet{Sa12}, particularly 
NGC~5927 and NGC~6528, have notably smaller $W^{\prime}$ values than would be expected for
their abundances and the linear calibration given in equation \ref{eq1}, necessitating the cubic calibration relation.
It is not at all clear why these metal-rich clusters differ from the location of old open clusters with similar metallicities,
although small sample sizes (e.g., 4 stars in NGC~5927 and 7 in NGC~6528) combined with significant differential
reddening may play a role \citep[see also the discussion in][]{Sa12} .  
Other calibrations such as those of \citet{AC04}, \citet{Car07} and \citet{WC09}, for 
example, that have employed old open clusters for the metal-rich part of the calibration, generally find linear 
relations analogous to equation \ref{eq1}.   We further note that if we use the clusters Melotte~66, 47~Tuc, 
NGC~288, 1904, 2298, 4590 and 7078 to establish a linear relationship between $W^{\prime}$ as measured 
here or
in \citet{Sa12}, and $W^{\prime}$ as given by \citet{Car07}, then the relationship is well defined with an {\it rms}
dispersion of only 0.09 \AA.  If we then use that relationship to convert the $W^{\prime}$ values for NGC~5927
and NGC~6528 given by \citet{Car07} to the $W^{\prime}$ scale used here, then the resulting values of 5.94
and 6.33 \AA\/, respectively, are much closer to the calibration defined by equation \ref{eq1}.  
As further support we note that the
recent calibration presented by \citet{SV15}, which is shown as the dot-dash line in the lower panel of Fig.\ \ref{fig3},
is also consistent with that adopted here.

\section{Results and Discussion}

\begin{figure}
\centering
\includegraphics[angle=-90.,width=0.46\textwidth]{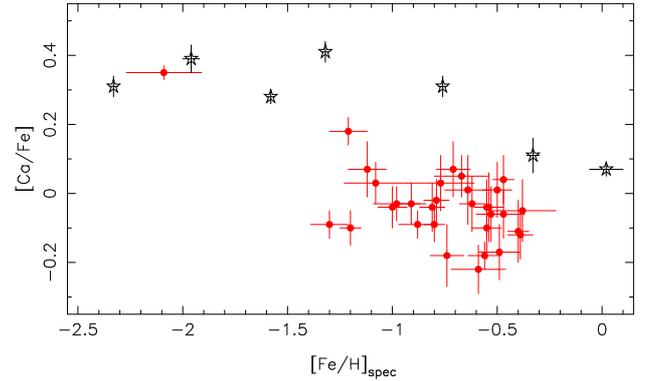}
\caption{The [Ca/Fe] abundance ratio as a function of [Fe/H] using values determined from high dispersion
spectroscopy analyses.  The black star symbols are for the Galactic calibration clusters while the red filled
circles are LMC disk red giants using data from \citet{vdS13}.  Only the LMC stars successfuly observed at the 
AAT from the \citet{Po08} sample are shown.
\label{fig4}}
\end{figure}

In Fig.\ \ref{fig4} we illustrate the issue under investigation by plotting [Ca/Fe] abundance ratios against
[Fe/H]$_{spec}$\footnote{The subscript {\it spec} is used to denote [Fe/H] determinations based on high
dispersion analyses.} for the Galactic calibration objects and for the 32 LMC disk red giants observed here that
pass the radial velocity and S/N checks. 
For the globular clusters the [Ca/Fe] values are taken from the high resolution study of \citet{Ca10} 
for NGC~7099, 1904, 288 and 47~Tuc, while that for NGC~2298 comes from \citet{McW92}, which is also based on
high dispersion spectroscopic data.  The [Ca/Fe] values for Melotte 66 and M67 come from
the high dispersion analyses of \citet{Ses08} and \citet{DY05}, respectively.  The abundances, abundance ratios, 
and their uncertainties for the LMC disk red giants are taken directly from \citet{vdS13}.  The LMC sample used
here has a median [Fe/H]$_{spec}$ of --0.69 and a median [Ca/Fe] of --0.04 dex.  At this [Fe/H]$_{spec}$ value 
the Galactic calibration objects typically have [Ca/Fe] ratios that are approximately 0.3 dex higher.

\begin{figure}
\centering
\includegraphics[angle=-90.,width=0.46\textwidth]{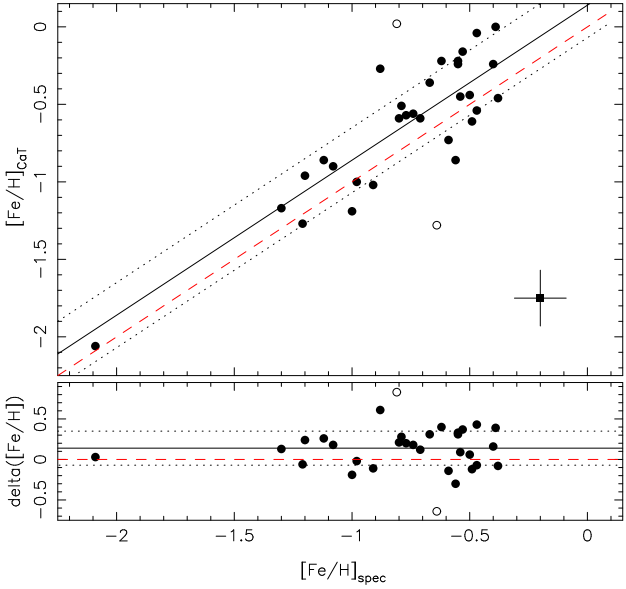}
\caption{The upper panel shows the relation between [Fe/H]$_{CaT}$ and [Fe/H]$_{spec}$ for the LMC
disk red giants.  The red dashed line is the 1:1 relation while the solid black line has a slope of unity but
an offset of 0.14 dex.  The dotted lines represent $\pm$1$\sigma$ from the solid line.  Two stars that deviate
by more than 2$\sigma$ are plotted as open circles.  Typical error bars in both quantities are shown the lower right
part of the plot.  The lower panel shows the difference ([Fe/H]$_{CaT}$ -- [Fe/H]$_{spec}$) as a function of 
[Fe/H]$_{spec}$.
\label{fig5}}
\end{figure}

In order to derive [Fe/H] values for the LMC stars using the calibration developed here, i.e., via equation \ref{eq1}, 
we proceed as follows.  The $V$ magnitudes of the LMC disk giants observed by \cite{Po08} are available 
on-line\footnote{http://cdsarc.u-strasbg.fr/viz-bin/Cat?J/A\%2bA/480/379}, while
\cite{AC00} list the mean magnitude of the red clump in this LMC disk field as $V$ = 19.26.  Adopting this value
as $V_{HB}$ then allows us to correct the measured $\Sigma$W for each LMC star to the equivalent $W^{\prime}$
value, which can then be used with equation \ref{eq1} to derive abundance values.  These are denoted by 
[Fe/H]$_{CaT}$.  Fig.\ \ref{fig5} shows the outcome of this process -- the upper panel plots  [Fe/H]$_{CaT}$ against
[Fe/H]$_{spec}$ while the lower panel shows the difference ([Fe/H]$_{CaT}$ -- [Fe/H]$_{spec}$) against 
[Fe/H]$_{spec}$ for the 32 LMC disk red giants.  
The two [Fe/H] determinations are clearly well-correlated but equally clearly a systematic 
offset is present.  After removing two stars that are more than 2$\sigma$ discrepant, the mean difference for the
remaining 30 stars is 0.14 dex, in the sense that the [Fe/H]$_{CaT}$ is more metal-rich, with a standard deviation
of 0.21 dex.   The standard deviation is consistent with that expected from the errors: $\sigma$[Fe/H]$_{spec}$ = 
0.11 dex \citep{vdS13} combined with a typical error in $\Sigma$W for the LMC stars of $\sim$0.3 \AA\/ and 
the calibration uncertainty of 0.06 dex.  Further, as is apparent in the lower panel of Fig.\ \ref{fig5}, aside from the 
offset between the two measures there is no evidence for any systematic dependence of the difference on 
[Fe/H]$_{spec}$, indicating that the abundance scales are consistent.  
This is not unexpected since both scales ultimately 
depend on high dispersion spectroscopic analyses but is reassuring nevertheless.  
The result here contrasts with that in 
\citet{vdS13} where, above [Fe/H]$_{spec}$ $\approx$ --0.5, there is an increasing difference between the [Fe/H]$_{spec}$ values and the [Fe/H]$_{CaT}$ values from \cite{AC05}.  The sense is that the [Fe/H]$_{CaT}$ values
become more metal-rich.

As for the systematic offset between the abundance values found here, it most likely has a straightforward 
explanation, one already anticipated to some extent by, for example, \citet{DH98} and \citet{Car07}.
Using the reddening for the LMC disk field given by \citet{vdS13},  E$(B-V)$ = 0.12 mag, and an LMC distance
modulus (m-M)$_{0}$ of 18.50 \citep{Piet13}, the adopted LMC $V_{HB}$ value of 19.26 corresponds to $M_{V}$ =
0.38 -- a value that is $\sim$0.4 mag brighter than the red clump stars in Melotte 66 or M67, for example, which have 
$M_{V}$ $\approx$ 0.8.  If a value of $V_{HB}$ value of 19.66 had been used instead for the 
LMC stars, then the
resulting metallicities would have been {\it lower} by 0.4 $\times$ 0.66 $\times$ 0.528 = 0.14 dex, precisely the
offset seen.  This reinforces the result that the abundance scale used in the calibration of the Ca~{\sc ii} line 
strengths is in excellent agreement with direct [Fe/H] determinations via high dispersion spectroscopy.  

It also suggests that the approach advocated by \citet{Car07,Car13}, where $M_{V}$ or $M_{I}$ is used in place of
$V_{HB}$, may be prefereable in situations where there is a large age range, or a dominant younger population, 
present, as is the case here.  In particular, as noted by \citet{Car07}, in systems like the disk of the LMC
where there has been extensive on-going star formation, the magnitude of the old core-helium burning stars, 
i.e., $V_{HB}$, is often not well defined as the red clump region of the CMD is dominated 
by more luminous younger stars.  In such situations, however, the distance modulus is often well established, 
e.g., via Cepheids, and thus the absolute magnitude $M_{V}$ is known more precisely than $V-V_{HB}$ 
\citep[see also][]{Po04}.
%\begin{figure}
%\centering
%\includegraphics[angle=-90.,width=0.46\textwidth]{fig6_paper.ps}
%\caption{The upper panel shows the relation between [Fe/H]$_{CaT}$ determined in this work and 
%[Fe/H]$_{CaT}$ listed in \citet{Po08} for the LMC disk red giants.  The red dashed line is the 1:1 relation while 
%the solid black line has a slope of unity but an offset of 0.04 dex.  The blue dotted lines represent $\pm$1$\sigma$ 
%from the solid line.  Three stars that deviate by more than 2$\sigma$ are plotted as open circles.  
%The lower panel shows the difference ([Fe/H]$_{CaT}$(this paper) -- [Fe/H]$_{CaT}$\citep{Po08} as a function of 
%[Fe/H]$_{CaT}$(this paper).
%\label{fig6}}
%\end{figure} 

\begin{figure}
\centering
\includegraphics[angle=-90.,width=0.46\textwidth]{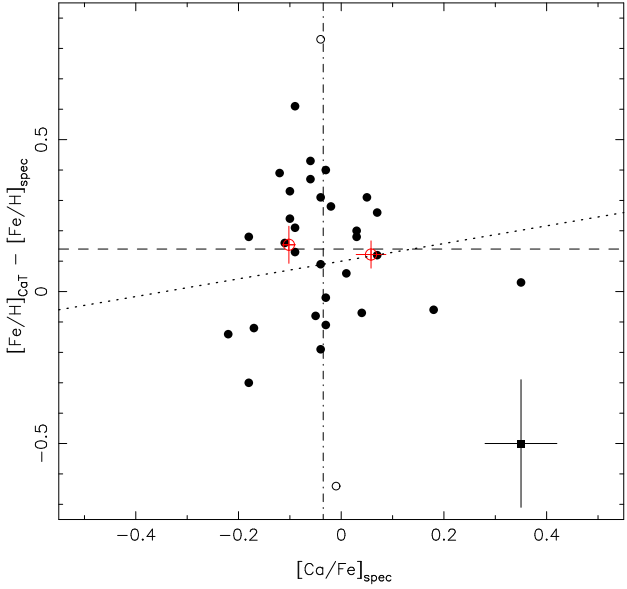}
\caption{The difference between [Fe/H]$_{CaT}$ and [Fe/H]$_{spec}$ for the LMC
disk red giants is plotted as a function of the [Ca/Fe] abundance ratio.  The horizonal dashed line shows the 
mean abundance difference for the sample, excluding the two discrepant stars (open circles), while the vertical 
dot-dash line shows the mean [Ca/Fe] value for the same stars.  
The red open circles represent the mean abundance difference for the stars with [Ca/Fe] less than, 
or greater than, the overall 
mean [Ca/Fe], respectively, plotted at the mean [Ca/Fe] for each sub-sample. 
The error bars on these points show the standard error of the means.
The standard deviation of the abundance differences and the mean error for the [Ca/Fe] determinations
are shown in the lower right of the figure.   The latter value is taken from the data of \citet{vdS13}.  
The dotted line is the relation between abundance difference
and [Ca/Fe] found from the data of \citet{Ba08} after an offset by 0.14 dex.
\label{fig7}}
\end{figure}

Turning now to the role of [Ca/Fe], we show in Fig.\ \ref{fig7} the relation between the difference 
[Fe/H]$_{CaT}$ -- [Fe/H]$_{spec}$ and [Ca/Fe] for the 32 LMC disk red giants.  If the [Ca/Fe] abundance ratio
plays a significant role in influencing the overall abundance estimated from the Ca~{\sc ii} line strengths, then
a positive correlation is expected in this diagram -- the abundance from the Ca~{\sc ii} triplet line strength and 
the Galactic calibration should underestimate [Fe/H]$_{spec}$ when the [Ca/Fe] value is low compared to the 
calibrators.
While the standard deviation of the abundance differences is sizeable, the data in 
Fig.\ \ref{fig7} demonstrate little compelling evidence for any significant dependence of the abundance difference
on the [Ca/Fe] abundance ratio.  In particular, if the data are separated into two sub-samples, one containing the
17 stars with [Ca/Fe] less than the overall mean, and one with the 13 stars where the [Ca/Fe] exceeds the overall 
mean, then the mean abundance differences for these two sub-samples do not differ.
%binning the data into three groups based on the [Ca/Fe] values
%reveals mean values that do not deviate substantially from the overall value.   Only for the bin containing the 
%4 stars with the lowest [Ca/Fe] abundance ratios does the mean for the bin differ from the overall mean in the 
%expected sense.   The effect, however, is barely more than 2$\sigma$.  The remaining stars with [Ca/Fe] $\geq$ 
%--0.15 show no indication of any significant dependence.  
While better signal-to-noise spectra at the Ca~{\sc ii} 
triplet for this sample of stars may have allowed tighter constraints, the data of Fig.\ \ref{fig7} indicate that the effect
on [Fe/H]$_{CaT}$ of [Ca/Fe] is relatively limited -- at most a $\sim$0.2 dex underestimate when [Ca/Fe] is 
substantially below the solar value.

\begin{figure}
\centering
\includegraphics[angle=-90.,width=0.46\textwidth]{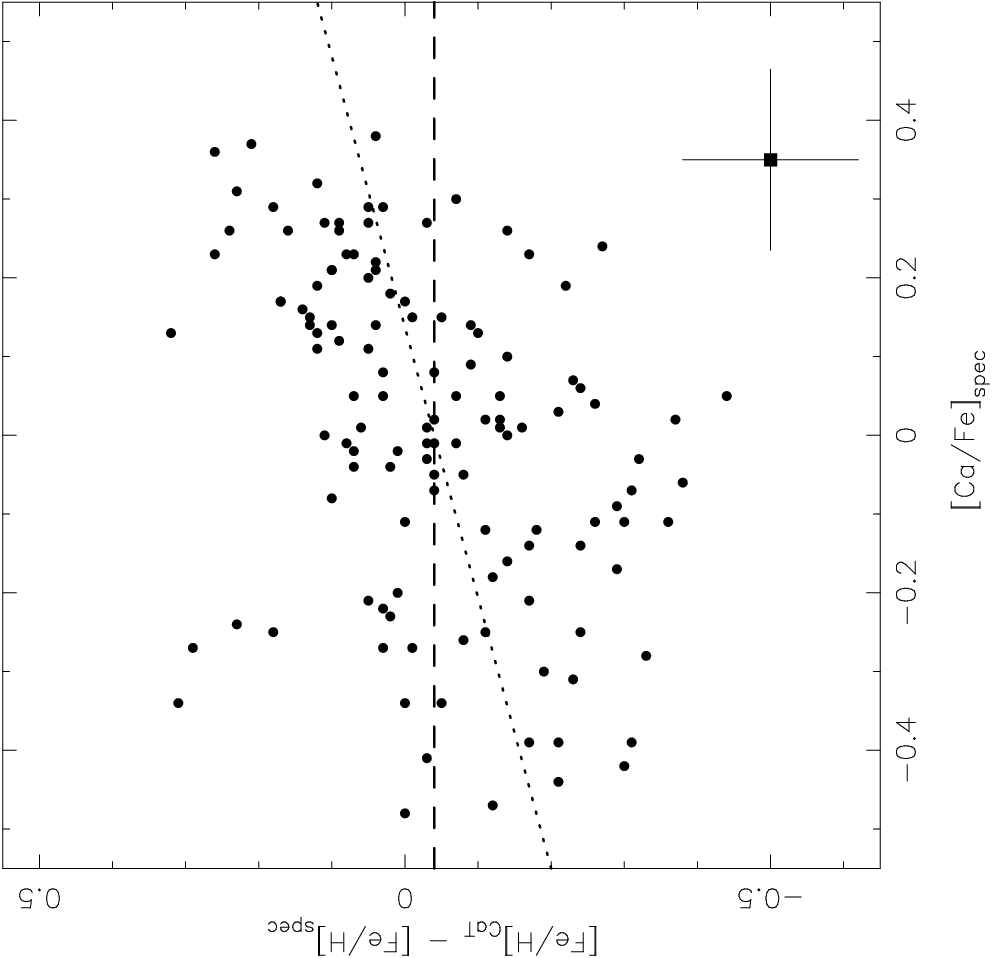}
\caption{The difference between [Fe/H]$_{CaT}$ and [Fe/H]$_{spec}$ for red giants in the Sculptor and
Fornax dwarf galaxies is plotted as a function of the [Ca/Fe] abundance ratio using data from
\citet{Ba08}.  As for Fig.\ \ref{fig7} the horizonal dashed line shows the 
mean abundance offset for the full sample.  %The red open circles represent the mean offset in the [Ca/Fe] bins
%outlined by the blue vertical dot-dash lines.  
The standard deviation of the abundance differences and the mean error for the [Ca/Fe] determinations
are shown in the lower right of the figure.  The dotted line is a least squares fit to the full sample.
\label{fig8}}
\end{figure}

In many respects the results presented here are a confirmation of those presented in \citet{Ba08} who also
found little indication of any substantial dependence of [Fe/H]$_{CaT}$ on [Ca/Fe].  
In their work \citet{Ba08} obtained 
moderate resolution spectra at the Ca~{\sc ii} triplet, as well as high dispersion spectroscopy, for a large sample
of red giants in the Sculptor and Fornax dwarf Spheroidal galaxies.  The high dispersion results were used to
directly calibrate the Ca~{\sc ii} triplet line strengths in terms of [Fe/H]$_{spec}$.  Fig.~\ref{fig8} shows their data
presented in the same way as for Fig.~\ref{fig7}.   Not surprisingly given the calibration process used, there is
a negligible offset between [Fe/H]$_{CaT}$ and [Fe/H]$_{spec}$: formally for the 125 stars in
the combined Sculptor and Fornax samples, the mean difference [Fe/H]$_{CaT}$ -- [Fe/H]$_{spec}$ is \mbox{--0.04}
with a standard deviation of 0.11 dex.  %Fig.~\ref{fig8} also shows the mean abundance difference in 4 bins of
%[Ca/Fe].  
Again there is little evidence for any substantial dependence on [Ca/Fe] -- as for Fig.\ \ref{fig7}, the change in 
[Fe/H]$_{CaT}$ -- [Fe/H]$_{spec}$ over the full range of  [Ca/Fe] observed is relatively small, of order 0.2 dex.
A least squares fit to the data points in Fig.\ \ref{fig8} yields the following equation:
\begin{equation}
\label{eq3}
([Fe/H]_{CaT} - [Fe/H]_{spec}) = 0.293\pm0.065 [Ca/Fe] - 0.038\pm0.014
\end{equation}
The correlation coefficient from the least squares fit is 0.377 indicating that the null hypothesis, namely that the
data points are uncorrelated, can be ruled out with a high degree of significance.  In other words, there is a genuine
dependence of the abundance difference on [Ca/Fe].  As regards the slope of the relation, applying a bootstrap
resampling process \citep[e.g.,][]{WJ12}
with 1000 trials to the 125 ($\Delta$[Fe/H], [Ca/Fe]) pairs shown in Fig.\ \ref{fig8}, yields a mean
slope of 0.276 dex/dex, entirely consistent with that in equation \ref{eq3}.   The relation from equation \ref{eq3} 
is also shown in Fig.\ \ref{fig7}, after allowing for the 0.14 dex abundance offset, and is clearly 
compatible with the LMC results derived here.   Indeed a slope of $\sim$0.3 lies within the inter-quartile range of
those derived from the data in Fig.\ \ref{fig7} in a similar bootstrap resampling analysis.

In summary, a change [Ca/Fe] from a typical Galactic value
of +0.3 to a value of --0.2 typical of those seen in the LMC and dwarf spheroidal galaxies will result,
assuming the use of Galactic calibrators, in an underestimate of the true [Fe/H] value by $\sim$0.15 dex
when using the Ca~{\sc ii} triplet method to estimate the overall abundance.  Since effects of this size are
frequently less than the random error in the abundance determinations via the Ca~{\sc ii} triplet method
for red giants in relatively distant dwarf galaxies \citep[e.g.,][]{RL13}, where the chemical evolution history
is generally unknown, the results here reaffirm that the Ca~{\sc ii} triplet approach can be used in the knowledge 
that the outcomes are not significantly affected by ignorance of the actual [Ca/Fe] abundance ratios.  

%Finally we note that the continuum opacity H$^{-}$ where the electrons come from the low ionization metals which
%include Mg and Fe as well as Ca.

\section*{Acknowledgements}

The author would like to acknowledge research support from the Australian Research Council through
Discovery Grant programs DP120101237 and DP150103294.  He is also grateful for the support 
received during an extended visit to the Institute for Astronomy, University of Edinburgh during which this
paper was largely written.  Thanks also to Ken Freeman, Russell Cannon and Elizabeth Maunder who conducted 
the 2010 December observations, and to Andrew Cole for providing a copy of the line profile fitting code (whose
heritage can be traced back to that of Armandroff \& Da Costa ``plus \c{c}a change, plus c'est la m\^{e}me chose").  
This research has made use of the WEBDA database, operated at the Institute for Astronomy of the University 
of Vienna, and of the VizieR catalogue access tool, CDS, Strasbourg, France. The original description of the 
VizieR service was published in A\&AS 143, 23.

\end{document}